\documentclass{article}

\usepackage{arxiv}
\usepackage[utf8]{inputenc} 
\usepackage[T1]{fontenc}    
\usepackage{hyperref}       
\usepackage{url}            
\usepackage{booktabs}       
\usepackage{amsfonts}       
\usepackage{nicefrac}       
\usepackage{microtype}      
\usepackage{lipsum}
\usepackage{graphicx}
\graphicspath{ {./images/} }
\usepackage{amsmath}
\usepackage{graphicx}
\usepackage{amsmath}
\usepackage{subcaption}   

\title{Leveraging Depth Maps and Attention Mechanisms for Enhanced Image Inpainting}

\author{
    Jin Hyun Park \\
    Dept. of Computer Science and Engineering \\
    Texas A\&M University \\
    College Station, TX, 77843  \\
    \texttt{jinhyun.park@tamu.edu}  \\
    \And
    Harine Choi \\
    Dept. of Multidisciplinary Engineering \\
    Texas A\&M University \\
    College Station, TX, 77843  \\
    \texttt{harinec@tamu.edu}  \\
    \And
    Praewa Pitiphat \\
    Dept. of Computer Science and Engineering \\
    Texas A\&M University \\
    College Station, TX, 77843  \\
    \texttt{riolu\_san@tamu.edu}  \\
}

\begin{document}
\maketitle

\begin{abstract}
Existing deep learning-based image inpainting methods typically rely on convolutional networks with RGB images to reconstruct images. 
However, relying exclusively on RGB images may neglect important depth information, which plays a critical role in understanding the spatial and structural context of a scene. 
Just as human vision leverages stereo cues to perceive depth, incorporating depth maps into the inpainting process can enhance the model's ability to reconstruct images with greater accuracy and contextual awareness.
In this paper, we propose a novel approach that incorporates both RGB and depth images for enhanced image inpainting. 
Our models employ a dual encoder architecture, where one encoder processes the RGB image and the other handles the depth image. 
The encoded features from both encoders are then fused in the decoder using an attention mechanism, effectively integrating the RGB and depth representations. 
We use two different masking strategies, line and square, to test the robustness of the model under different types of occlusions. 
To further analyze the effectiveness of our approach, we use Gradient-weighted Class Activation Mapping (Grad-CAM) visualizations to examine the regions of interest the model focuses on during inpainting.
We show that incorporating depth information alongside the RGB image significantly improves the reconstruction quality. 
Through both qualitative and quantitative comparisons, we demonstrate that the depth-integrated model outperforms the baseline, with attention mechanisms further enhancing inpainting performance, as evidenced by multiple evaluation metrics and visualization.
    
\end{abstract}
\section{Introduction}
Image inpainting, the task of reconstructing missing regions in an image, has seen significant progress due to the advances in deep learning techniques. Traditional methods primarily relied on convolutional neural networks (CNNs) \cite{lecun1989backpropagation} to inpaint images, often using the surrounding context to predict missing pixels. However, such methods can produce artifacts like color discrepancies and blurriness, especially in regions with complex structures or large missing areas. To address these limitations, we propose a novel approach that integrates both RGB and depth images with an attention mechanism \cite{bahdanau2014neural, vaswani2017attention} for RGB image inpainting, leveraging the spatial context provided by depth information to improve the accuracy of the reconstruction.
\par 
Recent advancements have demonstrated the effectiveness of incorporating attention mechanisms into CNNs to enhance their ability to focus on important features in the image \cite{dai2021coatnet, dosovitskiy2020image, liu2021swin}. Our approach builds on this idea by employing two different attention mechanisms: a simple attention mechanism and multi-head attention. The simple attention mechanism (i.e., single-head attention) allows the model to focus on relevant areas of the image, while the multi-head attention mechanism enables it to attend to multiple aspects of the image simultaneously, enhancing the model's ability to integrate both RGB and depth information.
\par 
The inspiration behind our approach is partially drawn from biological vision systems, where stereo images (i.e., paired vision from left and right eyes, thus leading to obtaining depth information) are used to understand objects and scenes in three dimensions. By mimicking this process, we seek to develop a more robust model that leverages depth information to aid in the inpainting of complex structures and textures, improving overall image quality.
\par 
To effectively evaluate the depth-integrated model, we employ two different masking strategies, line and square masks, ensuring a comprehensive test of the model's capabilities in handling different types of missing data. Our results show that the depth-integrated model significantly outperforms the baseline model, which uses only RGB images, across multiple evaluation metrics. We also utilize Grad-CAM \cite{selvaraju2020grad} to further analyze the regions of interest for the model and observe that attention mechanisms enable the model to better understand contextual and structural information compared to a simple encoder-decoder architecture. This analysis reveals how integrating depth improves the model's ability to focus on relevant features, enhancing its inpainting performance.
\par 
In summary, we propose dual encoder models where one encoder processes the RGB image and the other handles the depth image. These encoders work together to provide rich feature representations that are merged in the decoder using attention mechanisms. We demonstrate that incorporating depth information into the inpainting process not only improves the quality of image reconstruction but also makes the model more capable of handling irregular and complex missing regions.

\section{Background and Related Works}
Image inpainting has progressed significantly with deep learning. Early works such as Context Encoders \cite{pathak2016context} introduced an encoder-decoder architecture with adversarial training, showing that learning semantic representations could enable plausible image completions. However, these early methods often struggled with complex textures and large missing regions. Later, DeepFill \cite{yu2018generative} introduced attention mechanisms that allowed models to better utilize surrounding context, highlighting the importance of selectively focusing on informative features for inpainting. These ideas motivate our use of attention to intelligently fuse information from RGB and depth images.
\par
In parallel, using depth information alongside RGB has shown promise for improving scene understanding. Eigen et al. \cite{eigen2014depth} demonstrated that predicting depth maps from RGB improves recognition and reconstruction tasks. Extending this, Xiong et al. \cite{xiong2019depthfcn} proposed a fully convolutional network that uses RGB guidance to inpaint missing regions in depth images, suggesting that cross-modal cues significantly enhance inpainting quality. Similarly, Keaomanee et al. \cite{keaomanee2019rgbd} applied color-guided interpolation to refine depth inpainting near object boundaries, further emphasizing the complementary relationship between RGB and depth data. These results support our approach of leveraging depth information to assist RGB inpainting.
\par
Recent advances in attention mechanisms for image restoration provide further motivation for our model design. Wang et al. \cite{wang2019parallax} introduced a parallax attention module for stereo image super-resolution, showing that attending across spatial scales and modalities can improve reconstruction of fine details. Similarly, Ni and Cheng \cite{ni2022dualpath} proposed a Dual Path Cross-Scale Attention Network that integrates features along both encoder and decoder paths, demonstrating the benefit of multi-scale feature fusion. Inspired by these works, our model uses a dual-encoder structure with attention to merge RGB and depth features at different semantic levels.
\par
Furthermore, transformer-based models have recently achieved state-of-the-art results in large-hole image inpainting by explicitly modeling long-range dependencies. Li and Wang \cite{li2023putplus} optimized Transformers to balance global context with local refinements, while Chen et al. \cite{chen2023semanticaware} introduced semantic-aware memory modules to guide missing region completion. These findings suggest that attention-based architectures, especially Transformers, are particularly well-suited for capturing both global structure and fine textures, a key requirement when fusing RGB and depth modalities. Furthermore, Yan et al. \cite{yan2023penet} proposed PE-Net, demonstrating that explicit edge modeling with Transformers significantly improves restoration quality, especially for complex structures. This motivates our attention-based fusion approach, which aims to capture both semantic and geometric consistency between RGB and depth cues.
\par
Despite these developments, limited work has been done to explicitly combine depth maps with RGB images for inpainting. Inspired by the biological use of stereo vision for depth perception, we propose a dual-encoder model that leverages depth to guide inpainting. By combining RGB and depth encoders via attention mechanisms, our model generates more coherent and accurate reconstructions.

\section{Methods}
\subsection{Dataset and Pre-processing}
We used a subset of the NYU Depth Dataset V2 \cite{Silberman:ECCV12}, which includes only the labeled instances from the official NYU Depth Dataset V2 images. To optimize computational efficiency and performance, we resized the images from their original size of (480, 640) to (240, 320). Using the original scale would require significantly higher computational power due to the need for deeper encoders and decoders. We found that a shallow encoder-decoder approach with the (240, 320) resolution was sufficient to validate our results. Both RGB and depth images were normalized before processing. For the masked image, instead of providing the masked region to the models, we applied the mask and used the masked image as input. However, we observed that this approach led to undesirable performance when using random depth masks. This will be discussed further in the Discussion section.

\subsection{Masking}
\begin{figure}[htbp]
  \centering
  \begin{subfigure}[b]{0.45\textwidth}
    \centering
    \includegraphics[width=\linewidth]{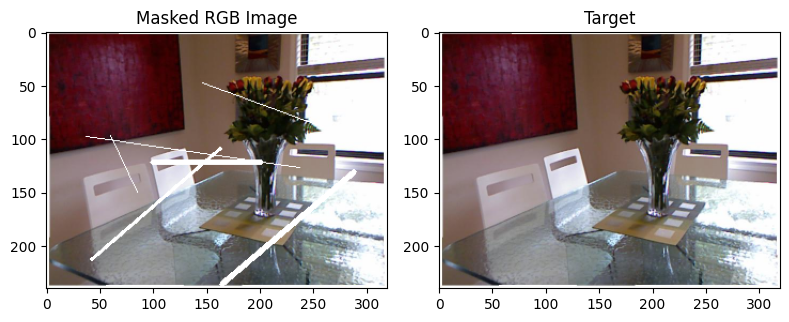}
  \end{subfigure}
  \hfill
  \begin{subfigure}[b]{0.45\textwidth}
    \centering
    \includegraphics[width=\linewidth]{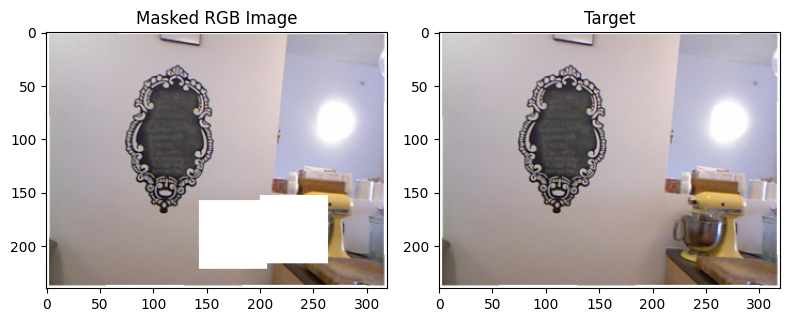}
  \end{subfigure}
  \caption{An example of line and square mask implementation. These masks are used for the baseline model, where the task is to reconstruct the RGB image from the RGB image. The same masking methods are used in both the baseline and depth-enhanced models.}
  \label{fig:line and square mask 1}
\end{figure}

\begin{figure}[htbp]
  \centering
  \begin{subfigure}[b]{0.45\textwidth}
    \centering
    \includegraphics[width=\linewidth]{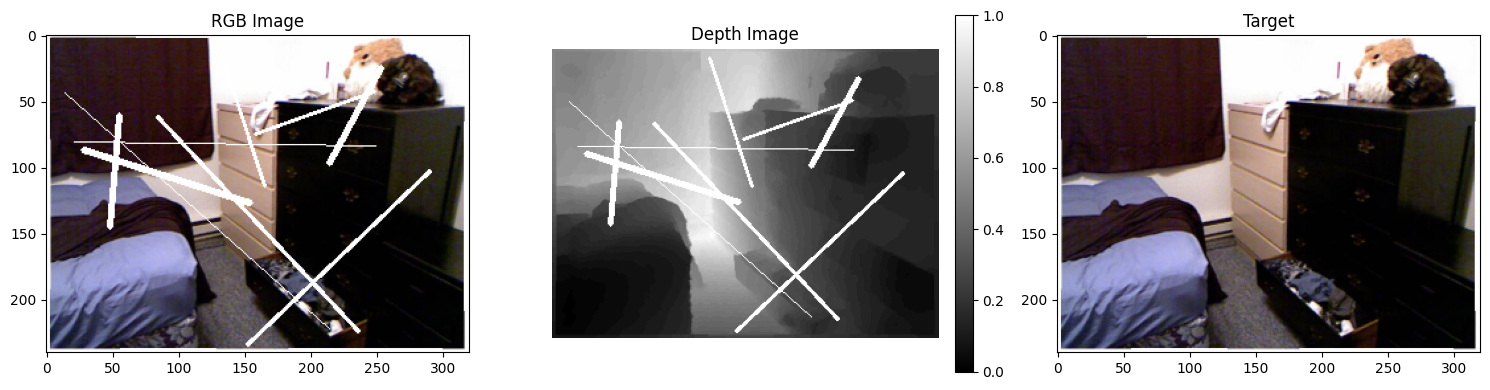}
  \end{subfigure}
  \hfill
  \begin{subfigure}[b]{0.45\textwidth}
    \centering
    \includegraphics[width=\linewidth]{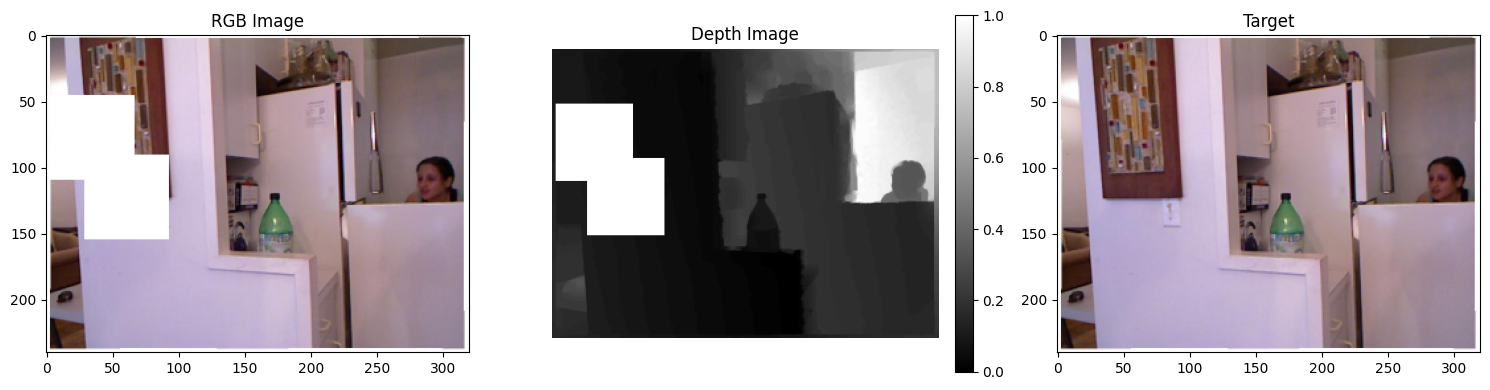}
  \end{subfigure}
  \caption{An example of line and square mask implementation. These masks are used for the depth-guided model, where the task is to reconstruct the RGB image using both the RGB image and the Depth image.}
  \label{fig:line and square mask 2}
\end{figure}

In our experiments, we used two types of masks: line mask and square mask. Both masks occlude certain areas in the images with white lines. The line mask consists of randomly drawn white lines, creating irregular, scattered occlusions that challenge the model to reconstruct the image. The square mask, on the other hand, occludes the image with randomly placed square blocks, drawn with white color. This structured occlusion is a more difficult task, requiring the model to better understand the spatial and depth relationships in the image and accurately infer missing content. We used the square mask to push the model to develop a deeper understanding of the depth input image. See Figure.~\ref{fig:line and square mask 1} for masked images trained with baseline model and Figure.~\ref{fig:line and square mask 2} for masked images trained with depth guided model. Note that the same mask and location were used for both the RGB image and the depth map.

\subsection{Evaluation Metrics}
We used PSNR, SSIM, LPIPS, and SSD to quantify the reconstruction accuracy between the reconstructed image and the original image. Below is a brief explanation of each metric:
\subsubsection*{Peak Signal-to-Noise Ratio (PSNR)}
    
    PSNR measures the power of the signal (the original image) relative to the power of the noise (the distortion introduced). A higher PSNR generally indicates better image quality. It's commonly used for evaluating lossy compression algorithms.  PSNR is typically expressed in decibels (dB).
        \begin{align*}
            PSNR &= 10\log_{10}\bigg(\frac{MAX_{I}^2}{MSE}\bigg)
        \end{align*}
    where $MAX_I$ is the maximum possible pixel value of the image. For example, if the image uses 8 bits per pixel, $(MAX_I)$ is 255.

\subsubsection*{Structural Similarity Index (SSIM)}

    SSIM goes beyond pixel-wise comparisons and considers the structural similarity between two images. It takes into account luminance, contrast, and structure.  SSIM values range from -1 to 1, with 1 indicating perfect similarity.
    \begin{align*}
        SSIM(x,y) &= \frac{(2\mu_x\mu_y + C_1)(2\sigma_{xy}+C_2)}{(\mu^2_x + \mu^2_y + C_1)(\sigma^2_x +\sigma^2_y +C_2)}
    \end{align*}
    where 
    \begin{itemize}
        \item $x$ and $y$ are image patches (small portions of the images). 
        \item $\mu_x$ and $\mu_y$ are the average pixel values of $x$ and $y$, respectively.
        \item $\sigma_x^2$ and $\sigma_y^2$ are the variances of $x$ and $y$, respectively.
        \item $\sigma_{xy}$ is the covariance of $x$ and $y$.
        \item $C_1$ and $C_2$ are small constants to stabilize the division.  Typically, $C_1 = (k_1L)^2$ and $C_2 = (k_2L)^2$, where $L$ is the dynamic range of pixel values (e.g., 255 for 8-bit images), and $(k_1)$ and $(k_2)$ are small constants (e.g., 0.01 and 0.03).
    \end{itemize}

\subsubsection*{Learned Perceptual Image Patch Similarity (LPIPS)}

    LPIPS measures perceptual similarity by comparing deep feature representations of images. It uses a pre-trained deep neural network (AlexNet \cite{krizhevsky2012imagenet} is used in our implementation) to extract features from image patches and calculates the distance between these feature vectors. LPIPS aims to align more closely with human visual perception than metrics like PSNR or SSIM. Lower LPIPS values indicate higher perceptual similarity. LPIPS doesn't have a single, simple equation like PSNR or SSIM. It involves the following steps:
    \begin{enumerate}
        \item Feature Extraction: Use a pre-trained convolutional neural network (CNN), denoted as $F$, to extract feature maps from the two input images.  Let $x$ and $y$ be the two input images. The CNN outputs feature representations $F(x)$ and $F(y)$.
        \item Normalization: The feature maps are normalized channel-wise.
        \item Distance Calculation: Calculate the squared Euclidean distance between the feature vectors at each spatial location.
        \item Averaging: Average the distances across all spatial locations and channels, possibly with channel-wise weights, to get the final LPIPS score.
    \end{enumerate}
    The general form can be represented as:
    \begin{align*}
        LPIPS(x,y) &= \sum_l \frac{1}{H_lW_l}\sum_{h,w}||w_l\odot(F_l(x)_{h,w}-F_l(y)_{h,w}||^2_2
    \end{align*}
    where
    \begin{itemize}
        \item $F_l(x)$ and $F_l(y)$ are the feature maps from the $l$-th layer of the network.
        \item $h$ and $w$ are the spatial indices.
        \item $H_l$ and $W_l$ are the height and width of the feature map at layer $l$.
        \item $w_l$ are channel-wise weights (learned during LPIPS training).
        \item $\odot$ denotes element-wise multiplication.
        \item $||\cdot||_2$ is the Euclidean norm.
    \end{itemize}

\subsubsection*{Sum of Squared Differences (SSD)}
    SSD is a very simple metric that calculates the sum of the squared differences between corresponding pixels in two images.  Lower SSD values indicate greater similarity.  SSD is very sensitive to even small changes in pixel values or image alignment.
    \begin{align*}
        SSD &= \sum^{H}_{i=1}\sum^{W}_{j=1}[I(i,j)-K(i,j)]^2
    \end{align*}
    where
    \begin{itemize}
        \item $H$ is the height of the image.
        \item $W$ is the width of the image.
        \item $I(i, j)$ is the pixel value at position $(i, j)$ in the original image.
        \item $K(i, j)$ is the pixel value at position $(i, j)$ in the reconstructed image.
    \end{itemize}

\subsection{Models}
    \subsubsection*{Baseline}
    The Baseline model utilizes a simple encoder-decoder, U-NET-like architecture \cite{ronneberger2015u} with convolutional layers. The encoder consists of multiple convolutional blocks, each followed by batch normalization and ReLU activation, and progressively downsamples the input image to extract hierarchical features. The decoder then upsamples the bottleneck feature map and combines it with skip connections from the encoder to reconstruct the image.
    \begin{itemize}
        \item Encoder: A series of convolutional blocks followed by max pooling layers to downsample the input. Skip connections are stored for later fusion in the decoder.
        \item Decoder: Upsamples the feature maps and concatenates them with corresponding skip connections from the encoder. A final convolution layer outputs the inpainted RGB image.
        \item Forward Pass: The model takes the corrupted RGB image as input, processes it through the encoder, and uses the decoder to generate the reconstructed image. The architecture does not incorporate any attention mechanisms. See SM Fig.~\ref{fig:baseline} for its detailed architecture. 
    \end{itemize}
    
    \subsubsection*{Depth Enhanced Simple Attention (DE-SHA)}
    The Depth Enhanced Simple Attention (DE-SHA) model extends the baseline architecture by incorporating a simple attention fusion mechanism to combine RGB and depth image features (simple attention works similarly to single-head attention; thus, we named our model DE-SHA). This model employs two separate encoders for processing RGB and depth images, respectively. The attention mechanism fuses the features by computing an attention map, which modulates the RGB features based on the depth features, allowing for more context-aware and depth-aware reconstruction.
    \begin{itemize}
        \item Encoder: Two separate encoders process the RGB and depth images. Each encoder performs convolutional operations followed by pooling to generate skip connections and bottleneck features.
        \item Attention Fusion: A simple attention fusion module combines the encoded RGB and depth features. The attention map is computed by passing concatenated RGB and depth features through convolutional layers, which generates a weight map to modulate the RGB features. Please see equation below for details. 
            \begin{enumerate}
                \item Concatenate the RGB and depth features:
                    \[
                    \mathbf{X}_{\text{Combined}} = \texttt{concat}(\mathbf{X}_{\text{RGB}}, \mathbf{X}_{\text{Depth}})
                    \]
                \item Compute the attention map:
                    \[
                    \mathbf{A} = \sigma \left( \texttt{conv}_2\left(\text{ReLU}\left(\texttt{conv}_1(\mathbf{X}_{\text{Combined}})\right)\right) \right)
                    \]
                    where \( \texttt{conv}_1 \) and \( \texttt{conv}_2 \) are convolution operations, and \( \sigma \) is the sigmoid activation.
                \item Apply the attention map: 
                    \[
                    \mathbf{X}_{\text{Fused}} = \mathbf{X}_{\text{RGB}} \cdot \mathbf{A} + \mathbf{X}_{\text{Depth}}
                    \]
                    where \( \cdot \) denotes element-wise multiplication.
            \end{enumerate}
        \item Decoder: The same as the baseline model. 
        \item Forward Pass: The encoder processes both the RGB and depth images to extract features, and the attention fusion mechanism is applied to combine the bottleneck features before passing them to the decoder for reconstruction. See SM Fig.~\ref{fig:simple attention} for its detailed architecture. 
    \end{itemize}

    \subsubsection*{Depth Enhanced Multi-head Attention (DE-MHA)}
    The Multi-head Attention model introduces self-attention mechanisms to fuse the encoded features of the RGB and depth images in the bottleneck stage. Instead of using a simple attention map, this model leverages multi-head self-attention to allow the model to focus on different parts of the input features simultaneously, improving the reconstruction performance by capturing complex relationships between the RGB and depth features.
    \begin{itemize}
        \item Encoder: Similar to the previous models, two separate encoders are used for the RGB and depth images. Each encoder applies a series of convolutions and pooling layers to generate hierarchical features.
        \item Multihead Self-Attention: After encoding, the bottleneck features from both the RGB and depth encoders are concatenated and passed through a multihead self-attention module. This allows the model to attend to different regions of the combined features simultaneously, enhancing the feature fusion process. Please see equation below for details.
            \begin{enumerate}
                \item Concatenate the RGB and depth features and linear projections:
                    The input feature maps from the RGB and depth encoders, \( \mathbf{X}_{\text{RGB}} \) and \( \mathbf{X}_{\text{Depth}} \), are concatenated to form the combined input feature map:
                    \[
                    \mathbf{X}_{\text{Combined}} = \texttt{concat}(\mathbf{X}_{\text{RGB}}, \mathbf{X}_{\text{Depth}})
                    \]
                    Then, we perform linear projections to compute the queries \( \mathbf{Q} \), keys \( \mathbf{K} \), and values \( \mathbf{V} \):
                    \[
                    \mathbf{Q} = \mathbf{X}_{\text{Combined}} \mathbf{W}_Q, \quad \mathbf{K} = \mathbf{X}_{\text{Combined}} \mathbf{W}_K, \quad \mathbf{V} = \mathbf{X}_{\text{Combined}} \mathbf{W}_V
                    \]
                    where \( \mathbf{W}_Q, \mathbf{W}_K, \mathbf{W}_V \) are learned weight matrices for the queries, keys, and values.
                \item Scaled dot-product attention and softmax normalization followed by a weighted sum of values:
                    \[
                    \mathbf{S} = \frac{\mathbf{Q} \mathbf{K}^T}{\sqrt{d_k}}, \hspace{1em}
                    \mathbf{A} = \texttt{softmax}(\mathbf{S}), \hspace{1em}
                    \mathbf{O} = \mathbf{A} \mathbf{V}
                    \]
                \item Multiple attention heads followed by a final linear transformation:
                    \[
                    \mathbf{O}_{\text{Multi-head}} = \texttt{concat}(\mathbf{O}_1, \mathbf{O}_2, \dots, \mathbf{O}_h), \hspace{1em}
                    \mathbf{O}_{\text{final}} = \mathbf{O}_{\text{Multi-head}} \mathbf{W}_O
                    \]
                    where \( h \) is the number of attention heads and \( \mathbf{W}_O \) is a learned weight matrix. We use \( h=4\) in our DE-MHA model.
            \end{enumerate}
        \item Decoder: The same as the baseline model. 
        \item Forward Pass: The RGB and depth images are processed through separate encoders, and their bottleneck features are fused using multihead self-attention. The resulting fused features are passed to the decoder, which reconstructs the RGB image. See SM Fig.~\ref{fig:multi-head attention} for its detailed architecture. 
    \end{itemize}
    
\subsection{Experimental Setup}
For the line masking experiment, we used a learning rate of \(1 \times 10^{-3} \) with the Adam optimizer. For the square masking experiment, we used the same learning rate of \( 1 \times 10^{-3} \) with the Adam optimizer, but with a weight decay of \( 1 \times 10^{-5} \).
The dataset is split into training, validation, and testing sets. For performance testing, we evaluated the model on the testing dataset using the model that achieved the best validation loss.
Regarding the encoder-decoder architecture, we used a 3-level encoder and a 3-level decoder. Each convolution layer uses a \( 3 \times 3 \) kernel with a padding of 1. The number of channels starts at 8 in the first level (3 to 8 for the RGB encoder and 1 to 8 for the depth encoder) and doubles with each subsequent level in the encoder. In the decoder, the number of channels is halved at each level. For each encoder, the images are down-sampled using 
\( 2 \times 2 \) max pooling with a stride of 2. In the decoder, the images are up-sampled by a factor of 2.

\section{Results}
\begin{table}[h!]
\centering
\footnotesize
\begin{tabular*}{\textwidth}{@{\extracolsep{\fill}}lcccc} 
\toprule
Method / Metrics           & SSD $\downarrow$        & PSNR $\uparrow$         & SSIM $\uparrow$          & LPIPS $\downarrow$       \\
\midrule
Baseline         & 0.09328                 & 1.56335                 & 1.00644                  & 0.41452                  \\
(Ours) DE-SHA      & \textbf{0.02182}  (-76.61\%)       & \textbf{1.92365}  (+23.05\%)      & \textbf{1.08205}   (+7.51\%)      & \textbf{0.08806}  (-78.76\%)       \\
(Ours) DE-MHA      & \textbf{0.01954}   (-79.05\%)    & \textbf{1.95208}  (+24.87\%)     & \textbf{1.09297}   (+8.60\%)     & \textbf{0.06690}   (-83.86\%)     \\
\bottomrule
\end{tabular*}
\vspace{1em}
\caption{Comparison of RGB inpainting methods using \textit{line} masks. Percentages represent the relative increase or decrease compared to the baseline performance.}
\label{tab:line-mask}
\end{table}
\begin{table}[h!]
\centering
\footnotesize
\begin{tabular*}{\textwidth}{@{\extracolsep{\fill}}lcccc} 
\toprule
Method / Metrics           & SSD $\downarrow$        & PSNR $\uparrow$         & SSIM $\uparrow$          & LPIPS $\downarrow$       \\
\midrule
Baseline         & 0.07968                 & 1.81590                 & 0.99252                  & 0.75372                  \\
(Ours) DE-SHA      & \textbf{0.07056}  (-11.45\%)        & \textbf{1.86784}  (+2.86\%)        & \textbf{1.02375}  (+3.15\%)        & \textbf{0.53271}  (-29.32\%)       \\
(Ours) DE-MHA      & \textbf{0.06813}  (-14.50\%)       & \textbf{1.89098}  (+4.13\%)       & \textbf{1.02495}  (+3.27\%)       & \textbf{0.52979}  (-29.71\%)      \\
\bottomrule
\end{tabular*}
\vspace{1em}
\caption{Comparison of RGB inpainting methods using \textit{square} masks. Percentages represent the relative increase or decrease compared to the baseline performance.}
\label{tab:square-mask}
\end{table}
We evaluate the performance of our model on two different masking strategies: line and square masks. The results are presented in Tables \ref{tab:line-mask} and \ref{tab:square-mask}, which show the quantitative comparisons using multiple evaluation metrics, including SSD, PSNR, SSIM, and LPIPS. 
In each table, we present the performance of each model using the ratio, which is defined by the following equation. The ratio is calculated as:
\[
\text{Ratio of inpainted to masked image using metric \textit{X}} = \frac{\sum_{i=1}^{n} X_i^{\text{inpainted}}}{\sum_{i=1}^{n} X_i^{\text{masked}}}
\]
where \( X = \{\text{SSD}, \text{PSNR}, \text{SSIM}, \text{LPIPS}\} \), \( X_i^{\text{inpainted}} \) represents the value of \( X \) for the \( i \)-th inpainted and the ground truth image, \( X_i^{\text{masked}} \) represents the value of \( X \) for the \( i \)-th masked and the ground truth image, and \( n \) is the total number of images in the test dataset.
\par 
In terms of performance, both DE-SHA and DE-MHA outperformed the baseline model across all metrics. For the line mask experiment, DE-MHA showed the most significant improvement, achieving a reduction in SSD by 79.05\% and LPIPS by 83.86\%. Similarly, there is an increase in PSNR by 24.87\% and SSIM by 8.60\%. Similarly, DE-SHA significantly improved, with SSD reduced by 76.61\% and LPIPS reduced by 78.76\%. PSNR increased by 23.05\% and SSIM increased by 7.51\%. The square mask experiment showed similar trends, with DE-MHA outperforming DE-SHA in all metrics, while both models significantly outperformed the baseline model.
\par
\begin{figure}[htbp]
  \centering
  \begin{minipage}{0.32\linewidth}
    \centering
    \includegraphics[width=\linewidth]{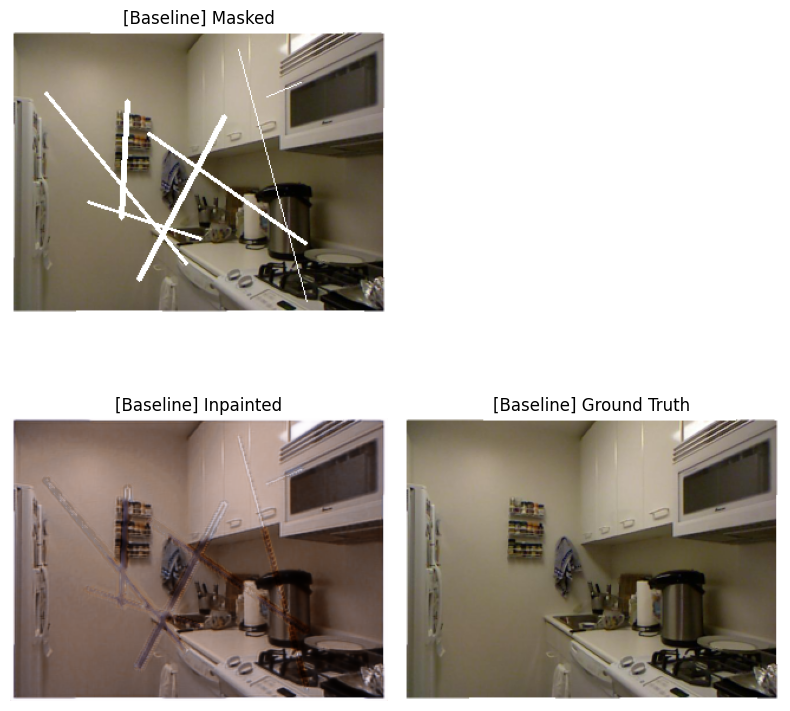}
  \end{minipage}\hspace{0.1cm} 
  \begin{minipage}{0.32\linewidth}
    \centering
    \includegraphics[width=\linewidth]{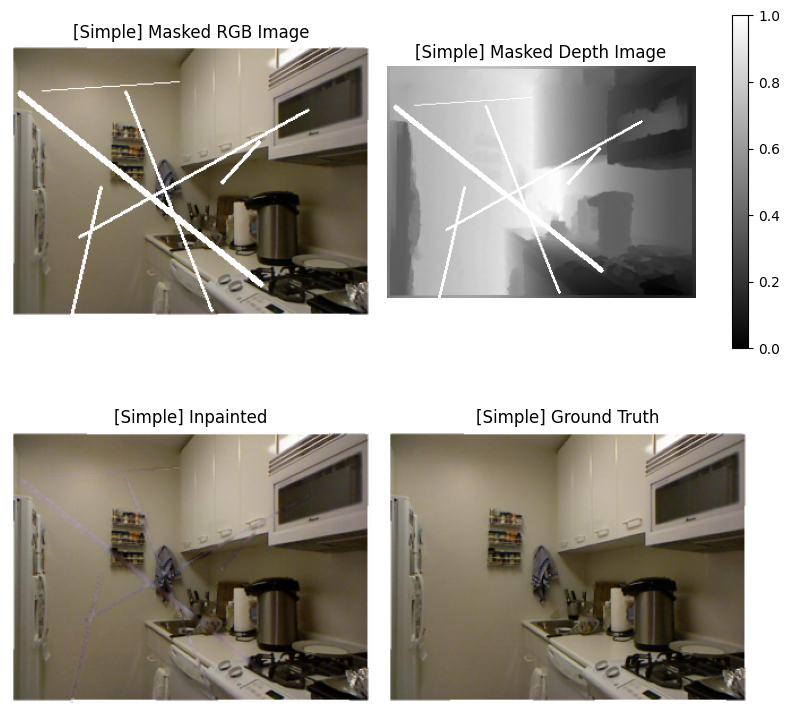}
  \end{minipage}\hspace{0.1cm}
  \begin{minipage}{0.32\linewidth}
    \centering
    \includegraphics[width=\linewidth]{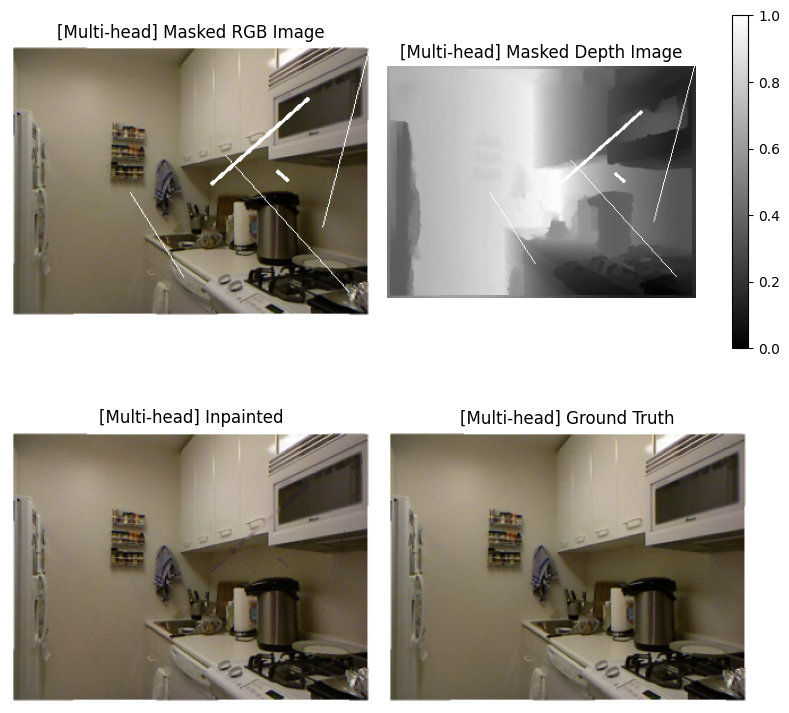}
  \end{minipage}
  \caption{Comparison of inpainting results using \textit{line} masks. The left four images show the baseline model’s inpainting, while the middle four images show the inpainting using the DE-SHA model, and the right four images show the inpainting using the DE-MHA model.}
  \label{fig: line 3}
\end{figure}

\begin{figure}[htbp]
  \centering
  \begin{minipage}{0.32\linewidth}
    \centering
    \includegraphics[width=\linewidth]{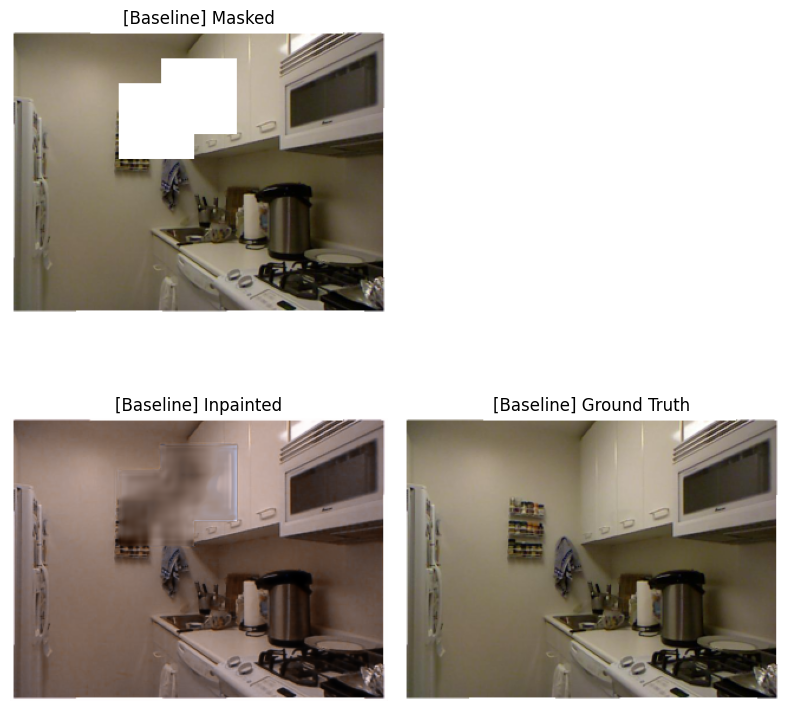}
  \end{minipage}\hspace{0.1cm} 
  \begin{minipage}{0.32\linewidth}
    \centering
    \includegraphics[width=\linewidth]{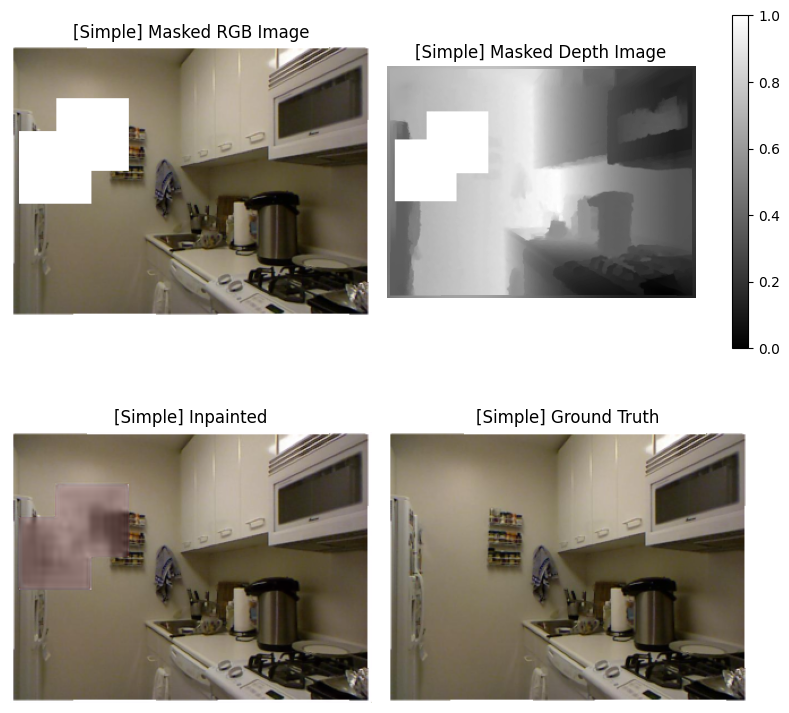}
  \end{minipage}\hspace{0.1cm}
  \begin{minipage}{0.32\linewidth}
    \centering
    \includegraphics[width=\linewidth]{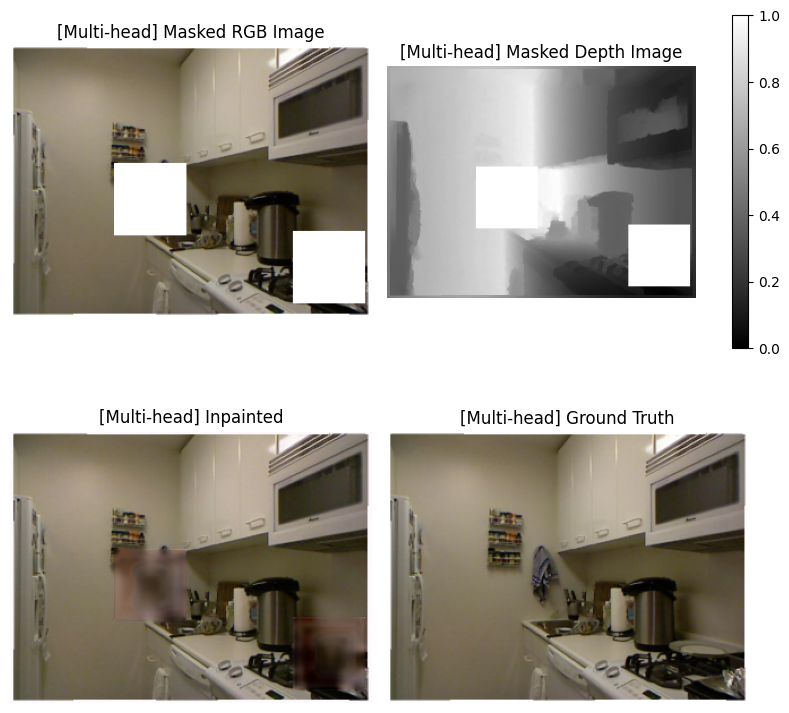}
  \end{minipage}
  \caption{Comparison of inpainting results using \textit{square} masks. The left four images show the baseline model’s inpainting, while the middle four images show the inpainting using the DE-SHA model, and the right four images show the inpainting using the DE-MHA model.}
  \label{fig: square 4}
\end{figure}

 
In the visual analysis, as shown in Figures \ref{fig: line 3} and \ref{fig: square 4}, the baseline model which relies only on RGB input reconstructs the image with noticeable artifacts, particularly a \textit{pinky} coloration in the reconstructed regions. In contrast, both DE-SHA and DE-MHA, which incorporate depth information, show much cleaner reconstructions with fewer artifacts. The absence of the pinky image in the inpainted results from DE-SHA and DE-MHA highlights the effectiveness of depth information in aiding RGB image reconstruction.
\par 
Our models, DE-SHA and DE-MHA, despite being shallow with only three levels in the encoder and decoder, still demonstrate superior performance, suggesting that the integration of depth information plays a critical role in improving image inpainting. The absence of artifacts like the pinky image in our models further confirms that depth information enhances the model's ability to reconstruct more accurate and realistic images. This strongly indicates that incorporating depth data significantly benefits overall image inpainting tasks, even with a relatively simple and shallow network architecture.


\section{Discussion}
Interesting insights emerge when analyzing Grad-CAM results with line and square masking. Grad-CAM visualizations are acquired from the last layer of the encoder, just before the decoder, allowing us to observe the regions of interest for the model during image inpainting.
\par
\begin{figure}[htbp]
  \centering
  \begin{minipage}{0.32\linewidth}
    \centering
    \includegraphics[width=\linewidth]{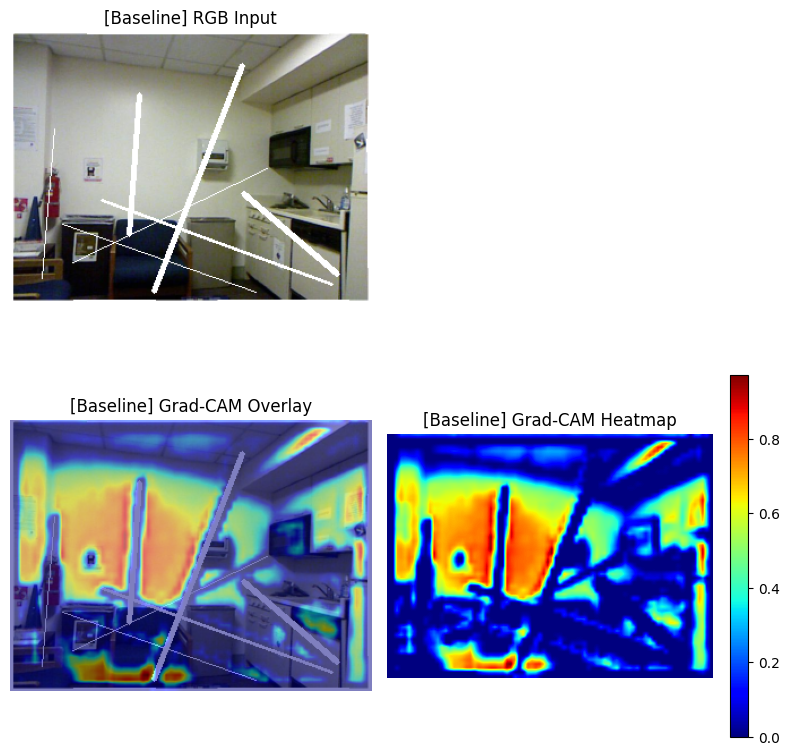}
  \end{minipage}\hspace{0.1cm} 
  \begin{minipage}{0.32\linewidth}
    \centering
    \includegraphics[width=\linewidth]{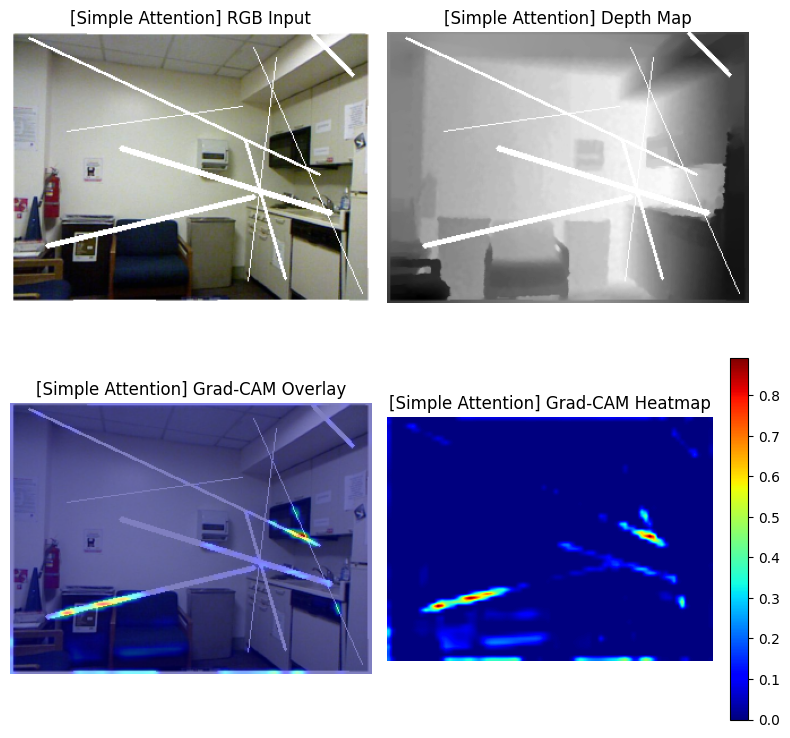}
  \end{minipage}\hspace{0.1cm}
  \begin{minipage}{0.32\linewidth}
    \centering
    \includegraphics[width=\linewidth]{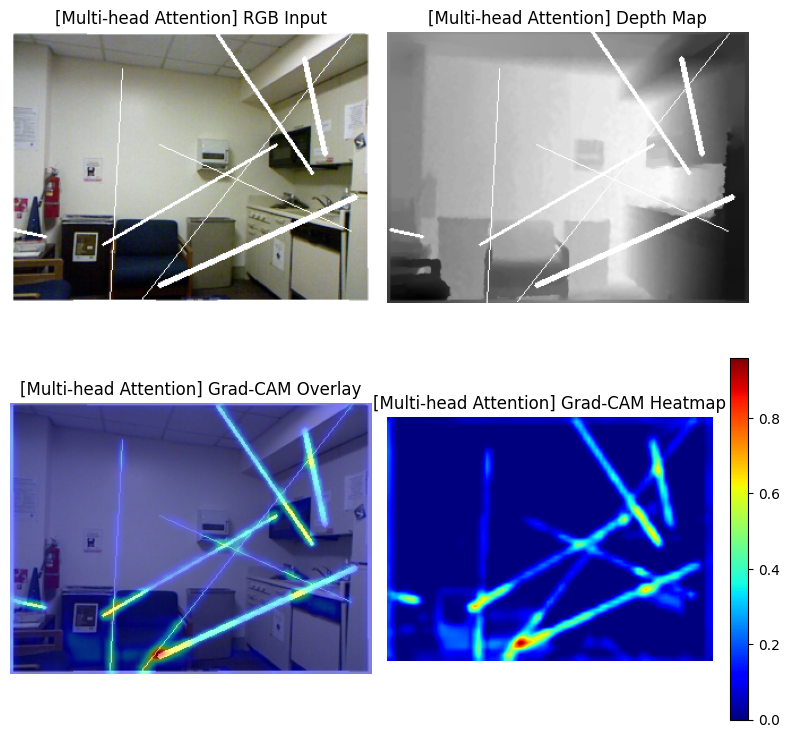}
  \end{minipage}
  \caption{Grad-CAM results for the base model, simple attention, and multi-head attention for line masking. Left: Baseline, Middle: DE-SHA, Right: DE-MHA.}
  \label{fig:3}
\end{figure}
For the line mask experiment, when only the RGB image is used for inpainting, we observe that the model tends to focus on the background rather than the lines themselves. The model appears to focus on the background as a way to capture the line segments. However, when the attention mechanism (either simple attention or multi-head attention) is used, the focus shifts to the lines themselves. A comparison between DE-SHA and DE-MHA reveals that the multi-head attention model is better at detecting the masked lines, resulting in improved inpainting performance. This behavior is intriguing because it resembles human perception: when presented with a masked RGB image, we tend to focus more on the missing line rather than the background.
\par
\begin{figure}[htbp]
  \centering
  \begin{minipage}{0.32\linewidth}
    \centering
    \includegraphics[width=\linewidth]{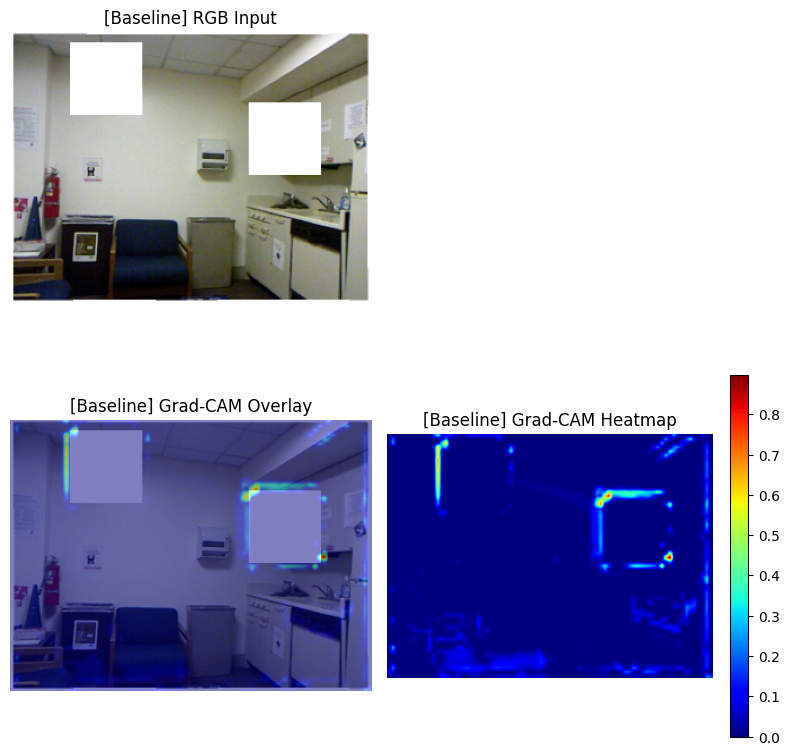}
  \end{minipage}\hspace{0.1cm} 
  \begin{minipage}{0.32\linewidth}
    \centering
    \includegraphics[width=\linewidth]{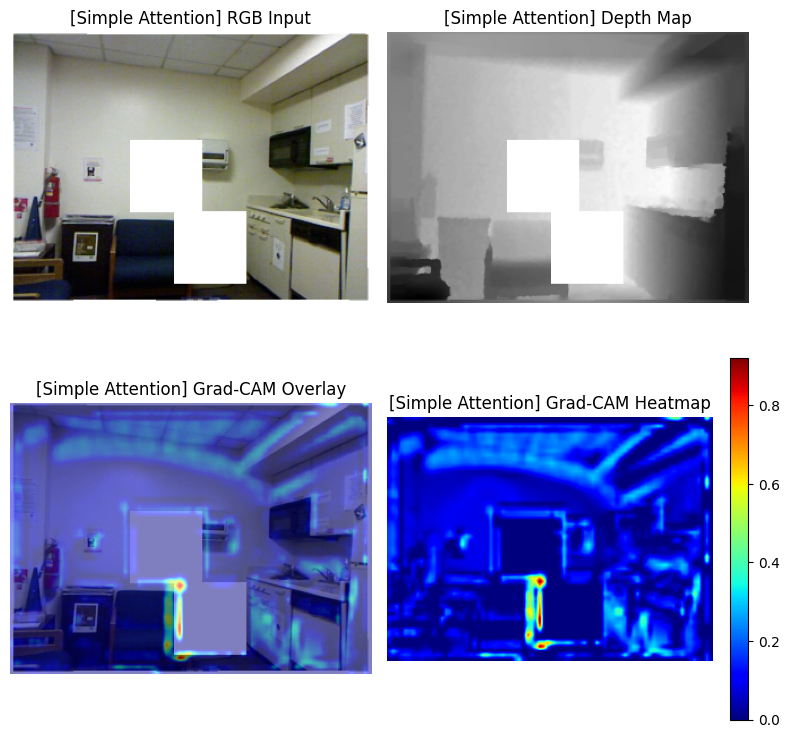}
  \end{minipage}\hspace{0.1cm}
  \begin{minipage}{0.32\linewidth}
    \centering
    \includegraphics[width=\linewidth]{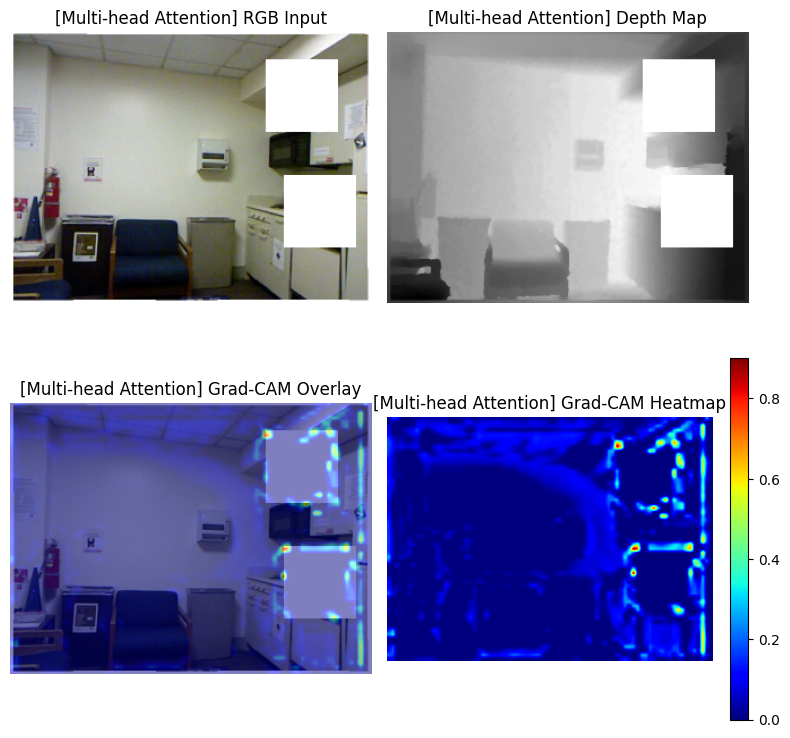}
  \end{minipage}
  \caption{Grad-CAM results for the base model, simple attention, and multi-head attention for square masking. Left: Baseline, Middle: DE-SHA, Right: DE-MHA}
  \label{fig:3}
\end{figure}
For the square mask experiment, attention models are better at capturing the overall structural information in the image. Moreover, the attention mechanisms focus more on the edges of the masked region than the baseline model. We hypothesize that this indicates a higher level of understanding of the image, where the model is more adept at identifying and reconstructing the missing structural details. This also explains why the depth-integrated model can reconstruct the image without the ``pinky'' artifacts.
\par 
One limitation of our approach is that we did not explicitly provide the masked region to the model; instead, we only provided the already masked image. As a result, the model first needs to infer the location of the masked region, which may have negatively affected the performance of the inpainting task. We believe that providing the masked region explicitly to the model could lead to improved performance.

\section{Limitations and Future Work}
There are several limitations in our work. The first limitation is that we did not explicitly provide the masked region to the model, which may have led to lower performance compared to models that explicitly receive the masked region. This is further supported by our Grad-CAM analysis, which shows that the model does not sufficiently focus on the masked region. Another limitation is our reliance on a CNN-based encoder-decoder architecture, rather than a Vision Transformer (ViT) model. Recent advances have demonstrated the effectiveness of ViTs in image-related downstream tasks, and several image inpainting models utilizing ViTs have shown promising results. We believe that replacing the CNN with a ViT could yield even better results when both RGB and depth information are given. However, our work still provides valuable insights by demonstrating that integrating depth information significantly enhances RGB image inpainting.
\par 
In terms of future work, the emergence of large language models (LLMs) presents an exciting opportunity for improving image inpainting tasks. One potential direction is to leverage LLMs to generate descriptive captions for each image. These captions could then be used in a multimodal model to aid in reconstructing masked regions. For example, providing the model with a generated caption from LLMs such as ``There are two square masks for the RGB and depth images in the same regions, and it is an image of a kitchen with the following objects: \{\textit{$object_{0}$, $object_{1}$, $\cdots$, $object_{N}$}\}. Your task is to fill in these masks.'' could help improve the inpainting results. As recent research has shown, integrating Vision-Language Models (VLMs) enhances performance in various image downstream tasks, and we believe this could further improve the inpainting task.

\section{Conclusion}
In this paper, we have proposed novel approaches to RGB image inpainting by integrating depth information. Our model leverages a dual encoder architecture, where one encoder processes the RGB image and the other handles the depth map, with both encoded features fused in the decoder using attention mechanisms. We showed that incorporating depth information significantly improves the image reconstruction task, as evidenced by both qualitative and quantitative evaluations.
\par
Through our experiments with two distinct masking strategies, line and square masks, we demonstrated that our depth-integrated model outperforms the baseline model in terms of reconstruction accuracy. The use of Grad-CAM further validated the effectiveness of the attention mechanisms, revealing that our model is better at focusing on the relevant structural and contextual information, such as the masked lines and edges, compared to the baseline model. Notably, the absence of artifacts such as the ``pinky'' image in the baseline inpainted results further emphasizes the beneficial impact of depth information.
\par
The results from both line and square masking tasks indicate that attention mechanisms allow the model to focus better on key regions (i.e., masked regions) of interest and lead to superior performance. This aligns with human-like perception, where we focus on the missing content in a masked image, while we partially use the surrounding background to infer the missing region.
\par
Overall, our findings suggest that depth information plays a crucial role in enhancing the quality of image inpainting and that attention mechanisms provide an effective way to better understand and reconstruct missing regions. Future work can explore further improvements in model architecture and apply this depth-enhanced approach to more complex inpainting tasks and real-world applications.

\section*{Reproducibility}
Our source code is available at \url{https://github.com/7201krap/CSCE748_Computational-Photography}

\bibliographystyle{unsrt}  
\bibliography{references}  

\begin{thebibliography}{10}

\bibitem{lecun1989backpropagation}
Yann LeCun, Bernhard Boser, John~S Denker, Donnie Henderson, Richard~E Howard, Wayne Hubbard, and Lawrence~D Jackel.
\newblock Backpropagation applied to handwritten zip code recognition.
\newblock {\em Neural computation}, 1(4):541--551, 1989.

\bibitem{bahdanau2014neural}
Dzmitry Bahdanau, Kyunghyun Cho, and Yoshua Bengio.
\newblock Neural machine translation by jointly learning to align and translate.
\newblock {\em arXiv preprint arXiv:1409.0473}, 2014.

\bibitem{vaswani2017attention}
Ashish Vaswani, Noam Shazeer, Niki Parmar, Jakob Uszkoreit, Llion Jones, Aidan~N Gomez, {\L}ukasz Kaiser, and Illia Polosukhin.
\newblock Attention is all you need.
\newblock {\em Advances in neural information processing systems}, 30, 2017.

\bibitem{dai2021coatnet}
Zihang Dai, Hanxiao Liu, Quoc~V Le, and Mingxing Tan.
\newblock Coatnet: Marrying convolution and attention for all data sizes.
\newblock {\em Advances in neural information processing systems}, 34:3965--3977, 2021.

\bibitem{dosovitskiy2020image}
Alexey Dosovitskiy, Lucas Beyer, Alexander Kolesnikov, Dirk Weissenborn, Xiaohua Zhai, Thomas Unterthiner, Mostafa Dehghani, Matthias Minderer, Georg Heigold, Sylvain Gelly, et~al.
\newblock An image is worth 16x16 words: Transformers for image recognition at scale.
\newblock {\em arXiv preprint arXiv:2010.11929}, 2020.

\bibitem{liu2021swin}
Ze~Liu, Yutong Lin, Yue Cao, Han Hu, Yixuan Wei, Zheng Zhang, Stephen Lin, and Baining Guo.
\newblock Swin transformer: Hierarchical vision transformer using shifted windows.
\newblock In {\em Proceedings of the IEEE/CVF international conference on computer vision}, pages 10012--10022, 2021.

\bibitem{selvaraju2020grad}
Ramprasaath~R Selvaraju, Michael Cogswell, Abhishek Das, Ramakrishna Vedantam, Devi Parikh, and Dhruv Batra.
\newblock Grad-cam: visual explanations from deep networks via gradient-based localization.
\newblock {\em International journal of computer vision}, 128:336--359, 2020.

\bibitem{pathak2016context}
Deepak Pathak, Philipp Krahenbuhl, Jeff Donahue, Trevor Darrell, and Alexei~A. Efros.
\newblock Context encoders: Feature learning by inpainting.
\newblock In {\em CVPR}, 2016.

\bibitem{yu2018generative}
Jiahui Yu, Zhe Lin, Jimei Yang, Xiaohui Shen, Xin Lu, and Thomas~S. Huang.
\newblock Generative image inpainting with contextual attention.
\newblock In {\em CVPR}, 2018.

\bibitem{eigen2014depth}
Christian~Puhrsch David~Eigen and Rob Fergus.
\newblock Depth map prediction from a single image using a multi-scale deep network.
\newblock In {\em NeurIPS}, 2014.

\bibitem{xiong2019depthfcn}
Rong Xiong, Guodong Liu, Yangyang Qu, and Yongsheng Ou.
\newblock Depth map inpainting using a fully convolutional network.
\newblock In {\em IEEE International Conference on Robotics and Biomimetics (ROBIO)}, 2019.

\bibitem{keaomanee2019rgbd}
Yossawee Keaomanee and Prakarnkiat Youngkong.
\newblock Rgb-d depth inpainting with color guide inverse distance weight.
\newblock In {\em International Conference on Information Technology (InCIT)}, 2019.

\bibitem{wang2019parallax}
Yue Wang, Jiayu Yang, Qingxiong Yang, Yuwen Xiong, Zhengfa Liang, Wei Zuo, and Lei Zhang.
\newblock Learning parallax attention for stereo image super-resolution.
\newblock In {\em CVPR}, 2019.

\bibitem{ni2022dualpath}
Yuanyuan Ni and Wengang Cheng.
\newblock Dual path cross-scale attention network for image inpainting.
\newblock In {\em IEEE International Conference on Image Processing (ICIP)}, 2022.

\bibitem{li2023putplus}
Zixuan Li and Yuan-Gen Wang.
\newblock Optimizing transformer for large-hole image inpainting.
\newblock In {\em IEEE International Conference on Image Processing (ICIP)}, 2023.

\bibitem{chen2023semanticaware}
Shiyu Chen, Wenxin Yu, Qi~Wang, Jun Gong, and Peng Chen.
\newblock Image inpainting with semantic-aware transformer.
\newblock In {\em ICASSP}, 2023.

\bibitem{yan2023penet}
Liang Yan, Chengyang Li, and Chengduan Wang.
\newblock Preserving edge with transformer for deep image inpainting.
\newblock In {\em IEEE International Conference on Information Technology, Big Data and Artificial Intelligence (ICIBA)}, 2023.

\bibitem{Silberman:ECCV12}
Pushmeet~Kohli Nathan~Silberman, Derek~Hoiem and Rob Fergus.
\newblock Indoor segmentation and support inference from rgbd images.
\newblock In {\em ECCV}, 2012.

\bibitem{krizhevsky2012imagenet}
Alex Krizhevsky, Ilya Sutskever, and Geoffrey~E Hinton.
\newblock Imagenet classification with deep convolutional neural networks.
\newblock {\em Advances in neural information processing systems}, 25, 2012.

\bibitem{ronneberger2015u}
Olaf Ronneberger, Philipp Fischer, and Thomas Brox.
\newblock U-net: Convolutional networks for biomedical image segmentation.
\newblock In {\em Medical image computing and computer-assisted intervention--MICCAI 2015: 18th international conference, Munich, Germany, October 5-9, 2015, proceedings, part III 18}, pages 234--241. Springer, 2015.

\end{thebibliography}

\newpage
\appendix
\section{Supplementary Materials (SM): Network Structures}

\begin{figure}[htbp]
  \centering
  \includegraphics[height=1.15\linewidth]{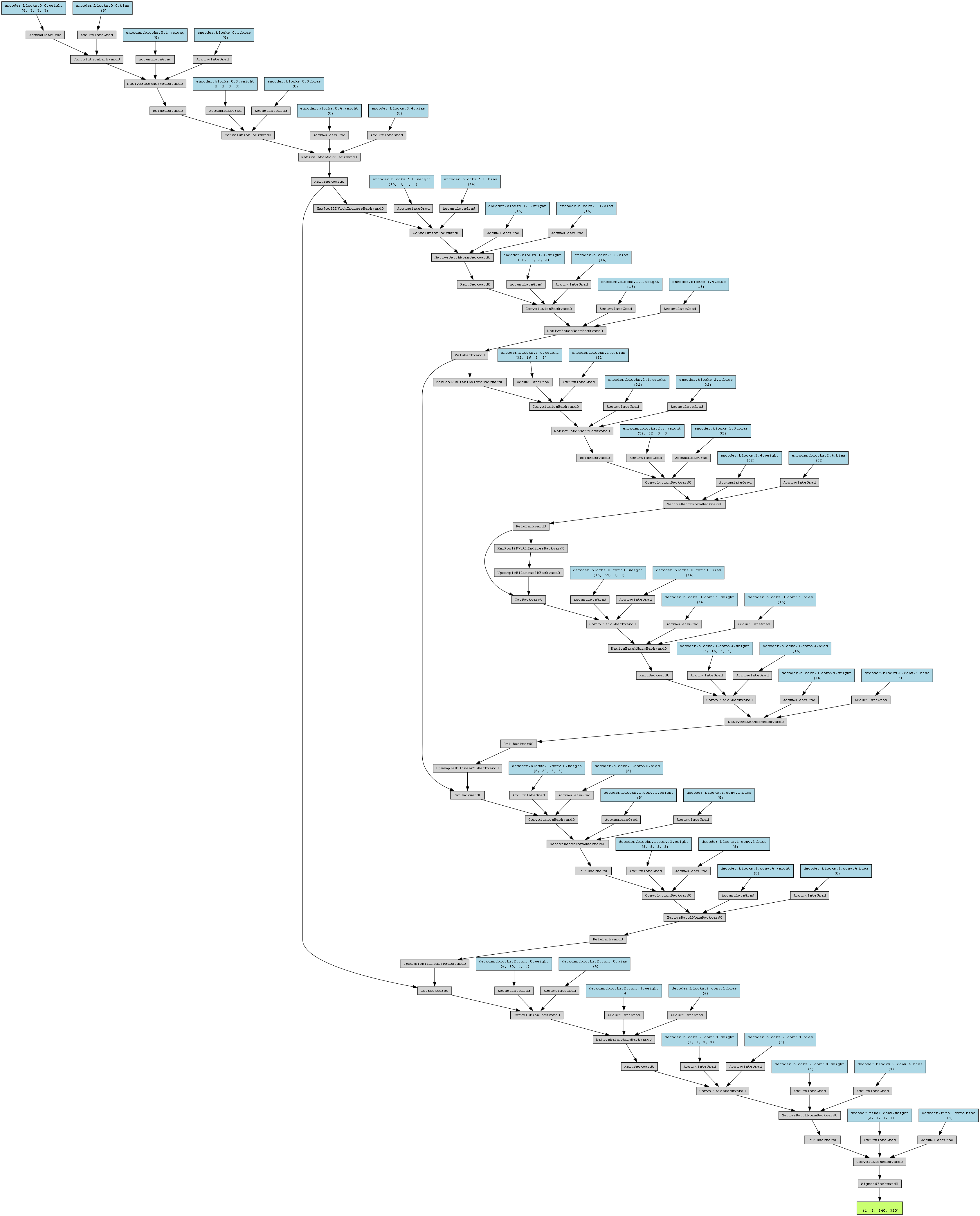}
  \caption{Block diagram of the baseline network.}
  \label{fig:baseline}
\end{figure}

\begin{figure}[htbp]
  \centering
  \includegraphics[height=1.15\linewidth]{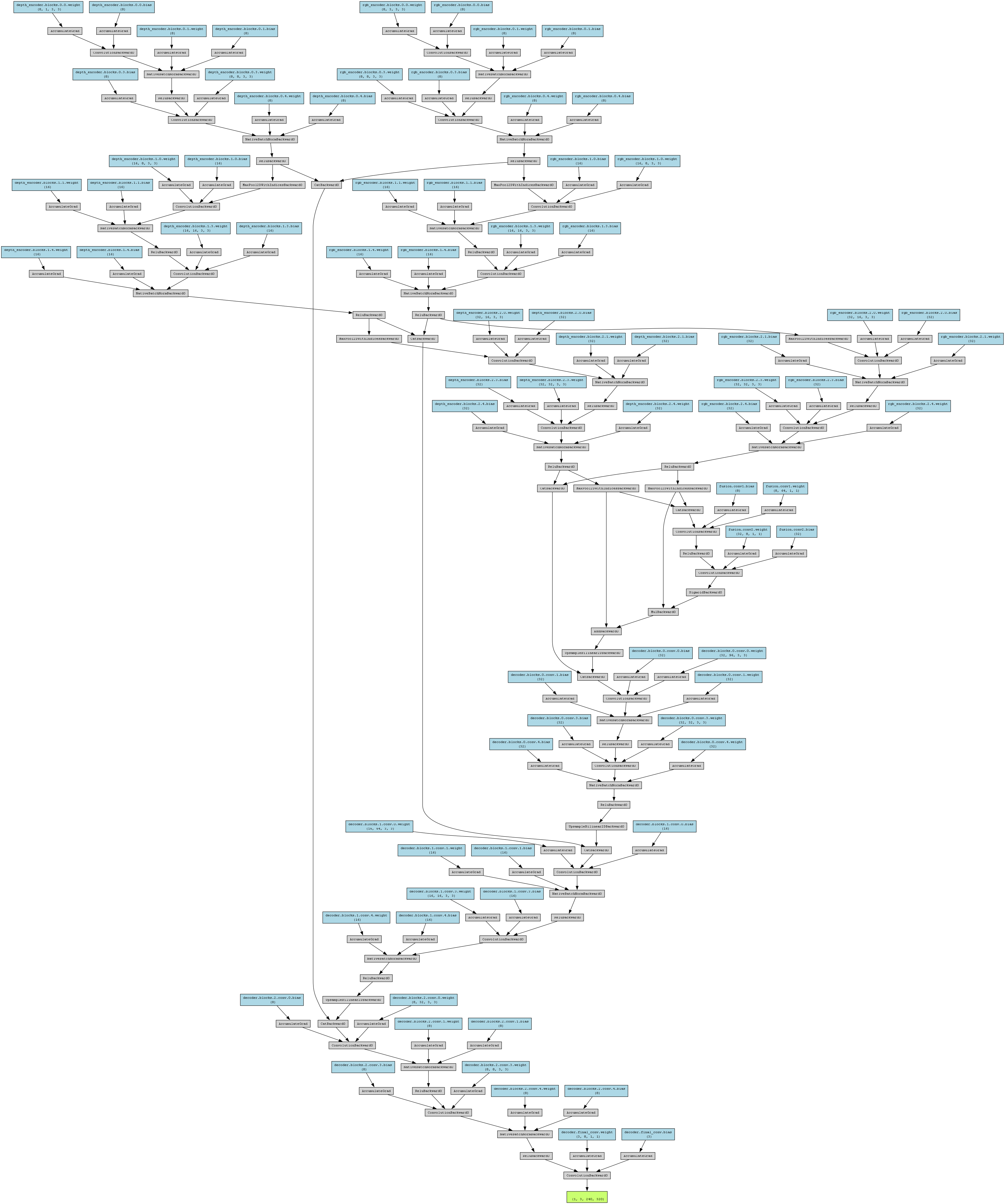}
  \caption{Block diagram of the depth‐enhanced simple attention network (DE-SHA).}
  \label{fig:simple attention}
\end{figure}

\begin{figure}[htbp]
  \centering
  \includegraphics[height=1.15\linewidth]{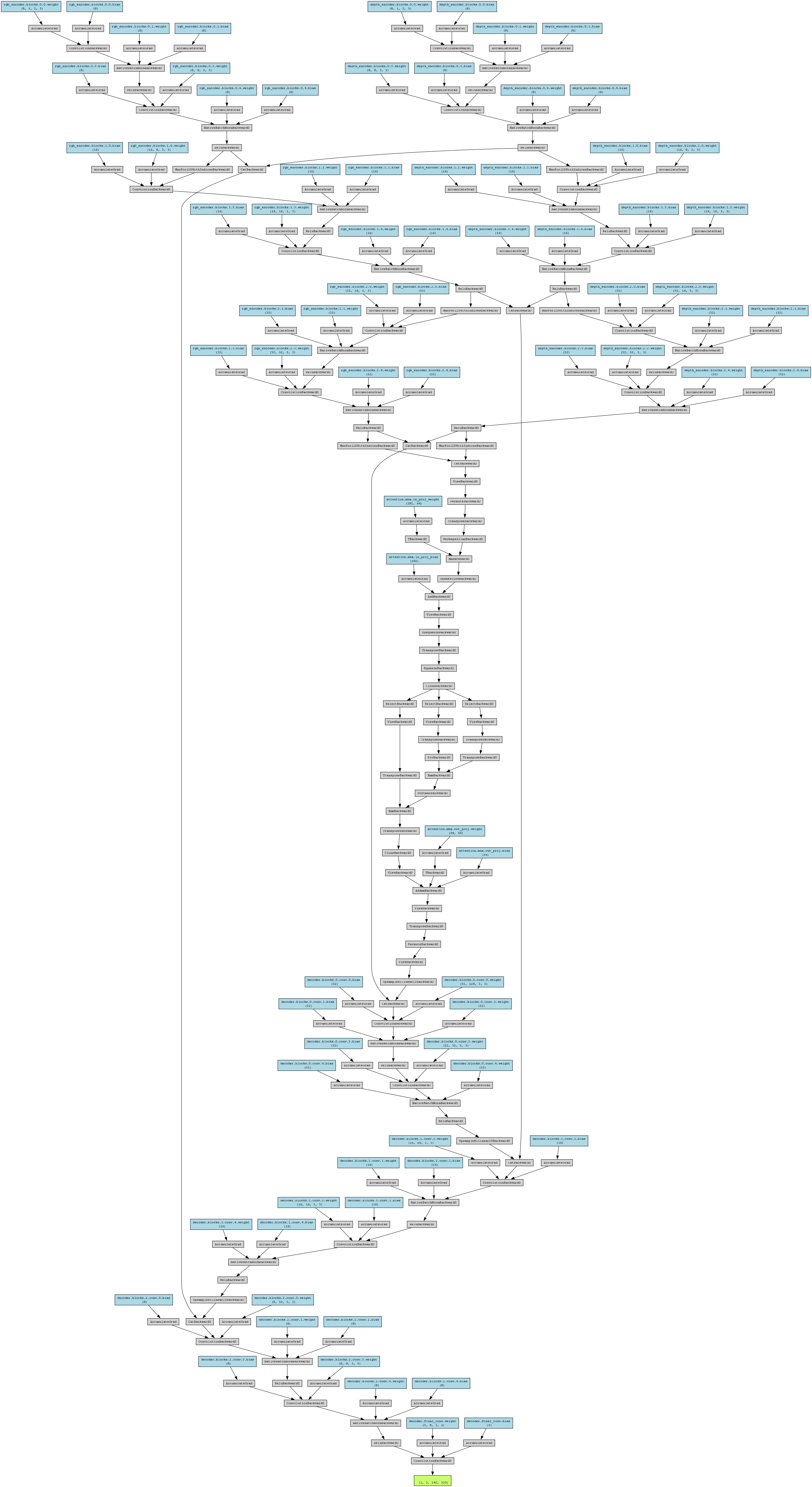}
  \caption{Block diagram of the depth‐enhanced multi-head attention network (DE-MHA).}
  \label{fig:multi-head attention}
\end{figure}

\end{document}